\documentclass[%
 reprint,
superscriptaddress,
 amsmath,amssymb,
 aps,
]{revtex4-2}

\usepackage{mathrsfs}
\usepackage{graphicx}
\usepackage{dcolumn}
\usepackage{bm}
\usepackage[paperwidth=200mm]{geometry}
\usepackage{xcolor}
\usepackage{amsmath}

\usepackage{fp}
\newcount\boxheight
\newcount\boxwidth
\usepackage{tabularx}
\newcommand\testaspect[1]{%
  \setbox0=\hbox{#1}%
  \boxheight=\ht0\relax%
  \boxwidth=\wd0\relax%
  \FPdiv\theaspect{\the\boxwidth}{\the\boxheight}%
  \copy0%
}

\usepackage{xr}

\usepackage{natbib}
\begin{document}


\title{Measuring Entanglement in Physical Networks}

\author{Cory Glover}
 \affiliation{Network Science Institute, Northeastern University, Boston, MA, USA}

\author{Albert-L\'{a}szl\'{o} Barab\'{a}si}
 \affiliation{Network Science Institute, Northeastern University, Boston, MA, USA}
 \affiliation{Department of Network and Data Science, Central European University, Budapest, Hungary}
 \affiliation{Department of Medicine, Brigham and Women’s Hospital, Harvard Medical School, Boston, MA, USA}

\begin{abstract}
The links of a physical network cannot cross, which often forces the network layout into non-optimal entangled states. Here we define a network fabric as a two-dimensional projection of a network and propose the average crossing number as a measure of network entanglement. We analytically derive the dependence of the crossing number on network density, average link length, degree heterogeneity, and community structure and show that the predictions accurately estimate the entanglement of both network models and of real physical networks. 
\end{abstract}

\maketitle

Many complex networks, from the brain \cite{towlson2013rich} to the network of atoms or molecules in materials have true physical manifestation, hence their nodes and links cannot cross each other.
While network science offers a series of tools to explore abstract networks like social networks or the World Wide Web, whose links are virtual and whose structure is fully encoded by the adjacency matrix $A_{ij}$, lately there is a growing interest in understanding physical networks, whose layout and properties are affected by the material nature of their nodes and links \cite{barabasi2023neuroscience}.
Indeed, volume exclusion and non-crossing conditions \cite{posfai2023impact,dehmamy2018structural,liu2021isotopy} can force such networks into non-optimal spatial layouts which they cannot escape, thereby creating entangled networks.
The resulting entanglement affects the network's elastic energy \cite{liu2021isotopy}, induces transitions in supercooled water \cite{neophytou2022topological} and affects the mechanical response of polymers \cite{grijpma1994chain}.

Here we investigate how the network embedding, defined by the detailed spatial layout of its nodes and links, and the network topology, captured by $A_{ij}$, affect network entanglement.
We begin by defining a \emph{network fabric} as a two-dimensional projection of a physical network. 
As a physical network can have multiple fabrics depending on the projection angle,
inspired by knot theory \cite{adams1994knot,fleming2005chord}, we propose the average crossing number (ACN) \cite{buck1999thickness,o2003energy} of a physical network as a measure of it's entanglement.
We derive analytically the dependence of entanglement on the network's density, link length, degree heterogeneity, and community structure.
Finally, we show that the developed analytical framework can predict changes in entanglement of both network models and real physical networks.

\begin{figure}[h!]
    \centering
    \includegraphics[width=\columnwidth]{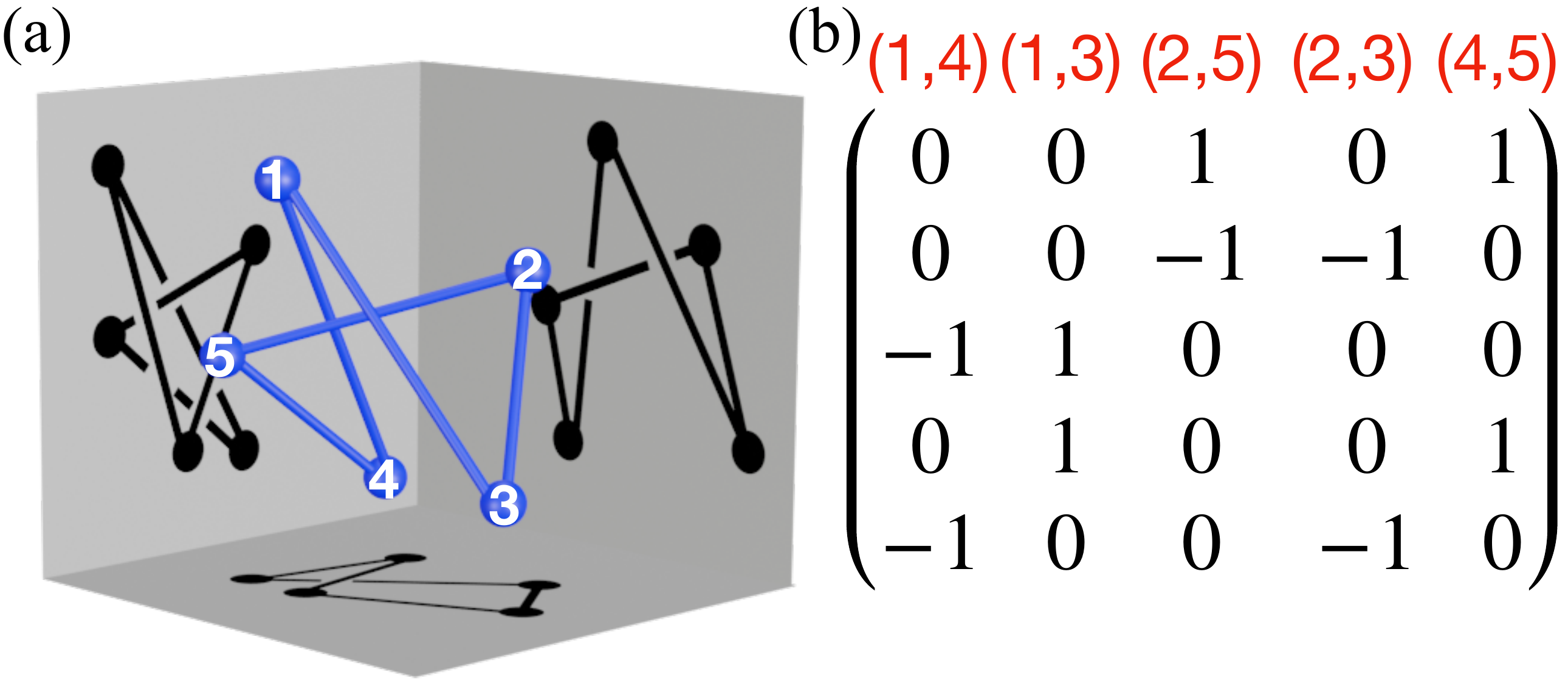}
    \caption{(a) Three fabrics of a simple five cycle. Each projection to a different plane results in a different fabric with different crossings. (b) The crossing matrix of the fabric in the left projection. Each row and column corresponds to a link in the physical network. To calculate the average crossing number, we average the number of crossings in each fabric. For the three fabrics shown in the figure, the five cycle has $\langle m\rangle\approx2.66$.}
    \label{fig:fabric}
\end{figure}

The fabric $f$  is a two-dimensional projection of a physical network such that each crossing point maintains the over and under crossing information (Fig.~\ref{fig:fabric}(a)).
The average crossing number (ACN) of a network is $\langle m\rangle=\frac{1}{4\pi r^2}\int_{S^2}m(f)dS$ \cite{buck1999thickness}, where $m(f)$ is the number of crossings in the fabric $f$, $dS$ is the area form of the sphere $S^2$ and $r$ is the radius of the sphere that holds the network.
In other words, $\langle m\rangle$ is the average number of link intersections over all possible fabrics (projections).

To estimate $\langle m\rangle$ for a given physical network, we take a projection and count the crossings in the resulting fabric using the crossing matrix, $R\in\mathbb{R}^{L\times L}$, whose entries are 1 if the row-link crosses over the column-link, $-1$ if the row-link crosses under the column-link, and 0 if the two links do not intersect (Fig.~\ref{fig:fabric}(b)).
We then average $m(f)$ over multiple fabrics to estimate the average crossing number.
We find that the ACN is self-averaging for large system sizes (see SI \ref{si:self-averaging}) and it correlates with the linking among loops (see SI \ref{si:gln}), suggesting that the crossings in a fabric directly informs our understanding of network entanglement.

For simplicity and without loss of generality, we focus on linear physical networks (LPNs), whose links are straight lines, and assume each link has infinitesimally small thickness.
This will allow us to focus on the effects of the network layout and topology on entanglement while avoiding the effects of volume.
Note that any non-linear physical network can be approximated with a LPN by adding ghost nodes along curved links, hence our results can be generalized to networks with curved links.

The network layout affects the probability that two links can cross,
which we can analytically calculate for random layouts (see SI \ref{si:max-crossings}) using Sylvester's four point problem \cite{pfiefer1989historical}.
However, real networks are not randomly embedded but follow an optimal layout, obtained by minimizing the network's total link length, which captures the system's elastic energy \cite{dehmamy2018structural,both2023accelerating,tutte1963draw,horvat2016spatial}.

The longer a link, the higher likelihood that it will cross other links, suggesting that the probability that each link pair will cross must scale with the average link length $\langle l\rangle$ (see SI \ref{si:max-crossings}).
Because real physical networks have widely different length scales, we normalize $\langle l\rangle$ by the average distance between two nodes $\langle l\rangle^*=\langle l\rangle/\langle d\rangle$.
As random layouts with higher $\langle l\rangle$ should have more crossings than optimal layouts with minimal $\langle l\rangle$, for random layouts $\langle l\rangle^*\rightarrow 1$ and for optimal layouts  $\langle l\rangle^*<<1$.
We then write $\langle m\rangle \sim \langle l\rangle^* m_{\max}$ where $m_{\max}$ is the maximum number of possible link pairs and $\langle l\rangle^*$ scales with the probability that the pair crosses.

In a LPN, if two links connect to the same node, they cannot cross elsewhere. 
Hence, $m_{\max}=\binom{L}{2}-\sum_{i\in V(G)}\binom{k_i}{2}$ \cite{peddada2021systematic}. 
Here the first term is the total number of link pairs and the second term removes the pairs which connect to the same node. We expand this bound to (see SI \ref{si:max-crossings}) \begin{equation}{m
_{\max}=\frac{L(L-1)}{2}-\frac{N}{2}\langle k^2\rangle+\frac{N}{2}\langle k\rangle},\label{eqn:m-max}\end{equation} where $\langle k\rangle$ and $\langle k^2\rangle$ are the moments of the degree distribution $P(k)$.

Real networks often present strong community structure in optimal energy layouts \cite{both2023accelerating,fortunato2010community}.
For spatially separated communities, links within the same community are more likely to cross each other than they are to cross links in different communities, suggesting that the presence of communities reduces $m_{\max}$.
To capture this reduction, consider a network with $C$ equal sized communities where each link in the network connects two nodes in distinct communities with probability $p$.
Then $(1-p)L$ links only cross links within their own community while $pL$ links can cross any link, reducing $m_{\max}$ to (see SI \ref{si:max-crossings})
\begin{equation}
        m_{\max}\approx\frac{(1-p)^2L^2}{2C}+2pL^2-\sum_i\binom{k_i}{2}.
        \label{eqn:m-max-comm}
\end{equation}

Combining (\ref{eqn:m-max}) and (\ref{eqn:m-max-comm}), we arrive at our key result, approximating the impact of the network layout and topology on LPN entanglement.
Defining $\langle m\rangle^*=\langle m\rangle/L^2$ as the normalized ACN, we obtain
\begin{widetext}

\begin{equation}
    \langle m\rangle^*\sim\langle l\rangle^*\Bigg(\frac{(1-p)^2}{C} + 2p- \frac{\langle k^2\rangle}{N\langle k\rangle^2}+\frac{1}{N\langle k\rangle}\Bigg).
    \label{eqn:avg-cross}
\end{equation}
\end{widetext}

This result highlights the combined impact of the network's embedding and topology on the network's entanglement.
Furthermore, it helps us understand the variables which control entanglement. 
To test the validity of (\ref{eqn:avg-cross}), we examine separately the role of the average link length, degree heterogeneity, and community structure.

\begin{figure}[h!]
    \centering
    \includegraphics[width=\columnwidth]{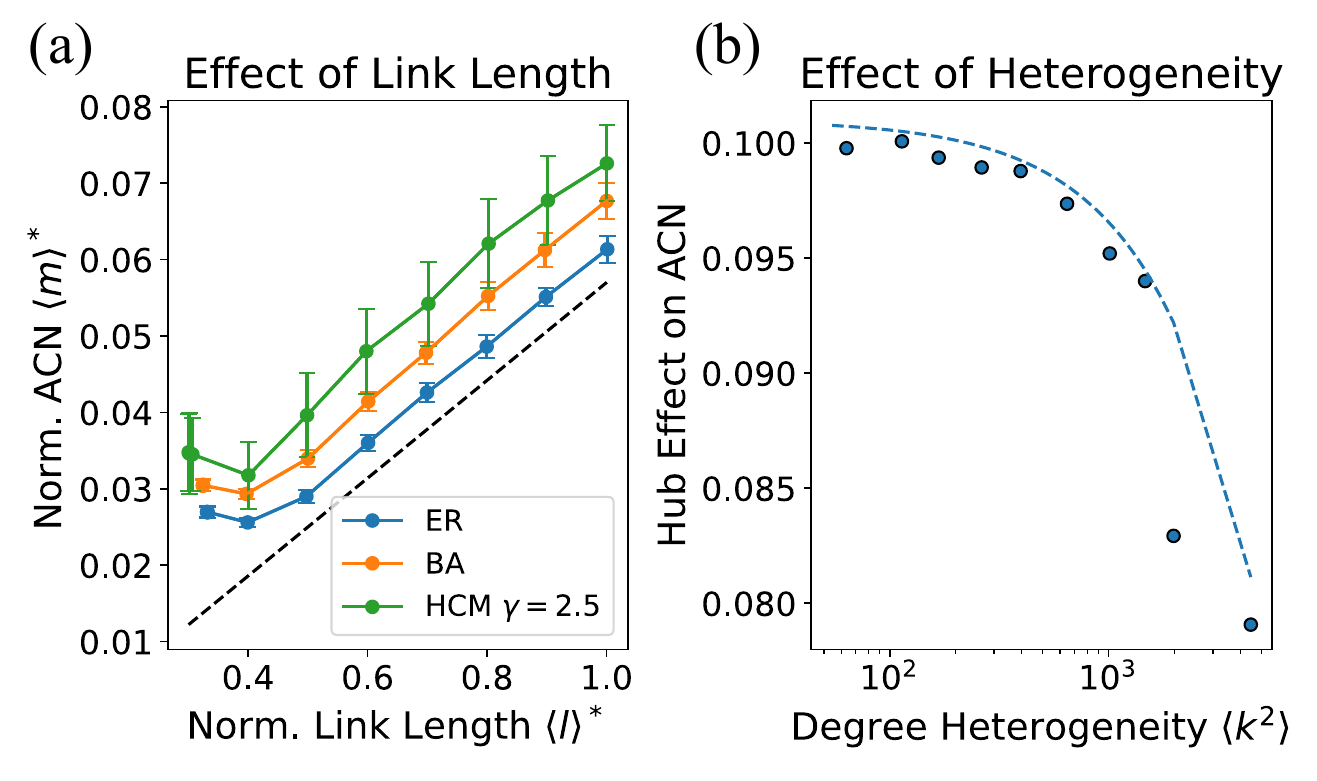}
    \caption{
    (a) The average link length for networks, generated by the ER, BA and configuration models with $N=10^3$ and $\langle k\rangle=6$, embedded using simulated annealing and compared with the normalized ACN. 
    The black dashed line corresponds to a linear fit $y=0.06x-.01$. 
    (b) We generate networks with $N=10^5$ and $P(k)\sim k^{-\gamma}$ using a hypercanonical configuration model with $\gamma=2.01$.
    The plot shows $\langle k^2\rangle$ versus $\frac{\langle m\rangle^*}{\langle l\rangle^*}$ and the dotted line is the analytical prediction (\ref{eqn:avg-cross}).
    }
    \label{fig:combined}
\end{figure}

\emph{Average Link Length} -- Networks with the same adjacency matrix $A_{ij}$ can have layouts with different $\langle l\rangle^*$.
To test the effect of $\langle l\rangle^*$ on the ACN, we generate different embeddings of the same network, each with a specified average link length $\langle l\rangle^*$, using simulated annealing (see SI \ref{si:mcmc}).
We measure the normalized ACN, $\langle m\rangle^*$, for each embedding.
As shown in Fig.~\ref{fig:combined}(a), we find a linear relationship between $\langle l\rangle^*$ and $\langle m\rangle^*$ for both ER, BA and configuration model networks, evidence that Eq.~\ref{eqn:avg-cross} correctly captures the ACN's dependence on $\langle l\rangle^*$ for networks with different layouts.

\emph{Degree Heterogeneity} -- A remarkable feature of Eq.~\ref{eqn:avg-cross} is its dependence on $\langle k^2\rangle$, indicating the unexpected role of degree heterogeneity and hubs in entanglement.
Indeed, while network robustness \cite{cohen2000resilience,caldarelli2007scale} and epidemic spreading \cite{boguna2003absence} are known to depend on $\langle k^2\rangle$, in physical networks the role of degree heterogeneity remains unknown.
Indeed, heterogeneity's role on LPNs emerges because links connected to a hub cannot cross.
Hence, hubs reduce the number of possible crossings.
In the extreme case of a star network, where all nodes are connected to a single hub, no links can cross and the ACN is zero.

Eq.~\ref{eqn:avg-cross} predicts the effect of heterogeneity for different network models.
Assume each model exhibits no community structure.
In regular or random regular networks ($k_i=k$), we have $\langle k^2\rangle=
\langle k\rangle^2$, hence as $N\rightarrow\infty$, $\langle m\rangle\sim \left(N\langle k\rangle\right)^2\langle l\rangle^*$.
In ER random networks, $\langle k\rangle=\langle k\rangle\left(1+\langle k\rangle\right)$,
thus, the leading term becomes $N(N-1)\langle k\rangle^2\langle l\rangle^*$.
In a scale-free network with degree exponent $\gamma$, we have $\langle k^2\rangle\sim N^{\frac{1}{\gamma-1}}$; hence, ACN scales as $\left(N\langle k\rangle\right)^2\langle l\rangle^*\left(1-N^{\frac{4-2\gamma}{\gamma-1}}+\frac{1}{N\langle k\rangle}\right)$.
When $\gamma<3$ the second moment diverges with the network size, resulting in a potentially significant reduction of the ACN.

To verify the role of degree heterogeneity, we generate networks with $N=10^5$ using a hypercanonical configuration model drawn from a power-law degree distribution \cite{voitalov2020weighted}.
For each network we measure $\langle m\rangle^*/\langle l\rangle^*$, finding that the dependence of $\langle m\rangle^*$ on $\langle k^2\rangle$ is well captured by Eq.~\ref{eqn:avg-cross} (see Fig.~\ref{fig:combined}(b)).

\begin{figure}[h!]
    \centering
    \includegraphics[width=\columnwidth]{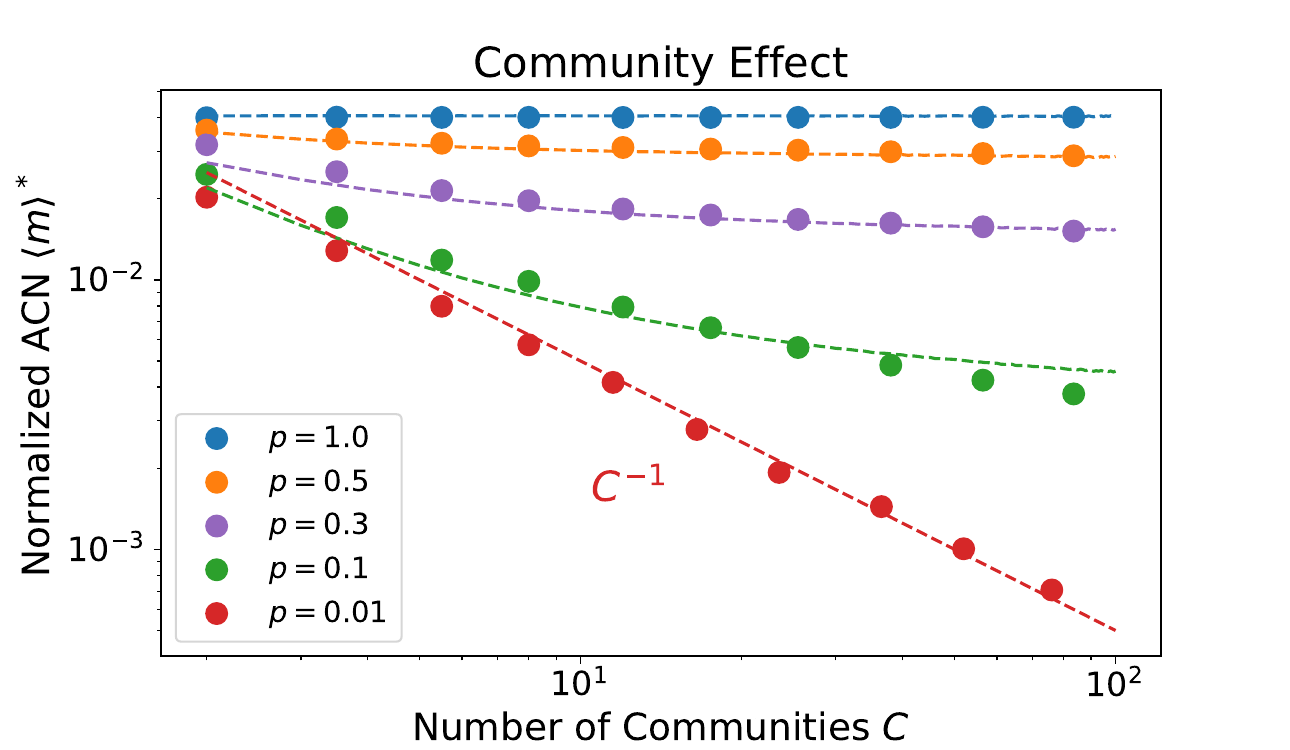}
    \caption{The normalized ACN for networks $(N=5000)$ with varying degree of inter-community connections ($p$). The dashed line corresponds to the prediction of curve Eq.~\ref{eqn:avg-cross}.}
    \label{fig:comm}
\end{figure}

\emph{Community Structure} -- Optimal network embeddings with community structure reduce entanglement by forcing links into distinct communities.
To test the role of community structure on the ACN, we generate networks with $C$ isolated, homogeneous communities.
Each link is then rewired to a random node in the network with probability $p$, representing an inter-community link.
As shown in Fig.~\ref{fig:comm}, we find that when $p\rightarrow 0$, $\langle m\rangle\sim C^{-1}$ as predicted in Eq.~\ref{eqn:avg-cross}.
For $p\rightarrow 1$, $C$ has no effect.
For intermediate $p$ values, we find that $\langle m\rangle\sim (1-p)^2C^{-1}+2p$ offers an excellent approximation, as predicted by Eq.~\ref{eqn:avg-cross} (see SI \ref{si:max-crossings}).

\begin{figure}[h!]
    \centering
    \includegraphics[width=\columnwidth]{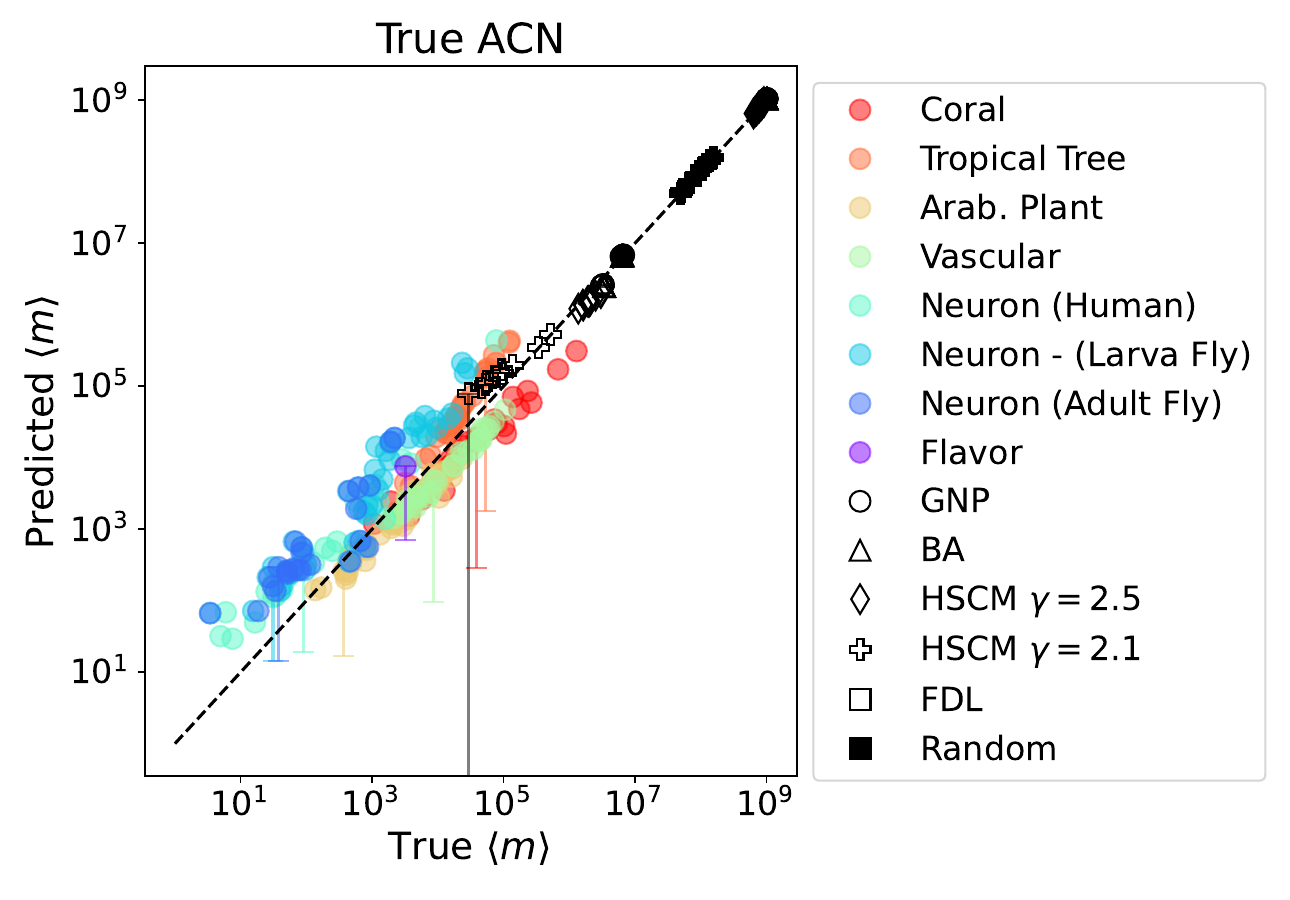}
\caption{The predicted ACN by Eq.~\ref{eqn:avg-cross} (vertical axis) for various real datasets, the flavor network, a BA network, an ER network, and two hypercanonical configuration models versus the emprically measured ACN (horizontal axis). Each synthetic network has 1000 nodes with average degree $\langle k\rangle=6$. The black dashed line is the diagonal. For each dataset we compute the ACN for $p=0$ and $p=1$, represented by the error bar (one for each dataset).
    For datasets, the data point is for $p=1$.}
    \label{fig:flavor}
\end{figure}

Taken together, we find that both degree heterogeneity and communities reduce network entanglement.
To test the accuracy of Eq.~\ref{eqn:avg-cross} as a whole, we generated networks using a variety of network models exhibiting varying levels of degree heterogeneity and community structure.
For each network, we predict the ACN using (\ref{eqn:avg-cross}) and calculate the true ACN numerically.
We embed each network with an optimal non-crossing layout offered by an accelerated force-directed layout \cite{both2023accelerating} and FUEL algorithm \cite{dehmamy2018structural} as well as in a random layout.
We estimate $C$ with the number of connected components.
Each network's ACN is then predicted for $p=0$ and $p=1$, obtaining a range of possible estimated ACN values.
For each network model, the prediction (\ref{eqn:avg-cross}) is close to the true ACN for both optimal and random layouts (Fig.~\ref{fig:flavor}).

Finally, we repeat the above procedure for real physical networks (see SI \ref{si:data}) and for the flavor network, a network with high heterogeneity and strong community structure, using a previously published optimal layout \cite{both2023accelerating}.
The number of communities $C$ is estimated by maximizing the modularity of the network.

Again, we find a good agreement between the predicted and true ACN (Fig.~\ref{fig:flavor}).
Similar results hold for the normalized ACN (see SI \ref{si:normalized-acn}).
For neurons, tropical trees, and the flavor network, the community structure must be accounted for to obtain a more accurate estimate.
However, the other networks' ACN is not affected by the community structure.

Previous work estimated network entanglement using the graph linking number (GLN) \cite{liu2021isotopy}, whic correlates with the ACN.
Yet, the GLN has an exceptionally high computational complexity and it is confined to networks with loops.
Indeed, for an arbitrary network, the GLN has complexity $O(N_L^2\langle c\rangle^2)$, where $N_L$ is the number of loops and $\langle c\rangle$ is the average loop size.
Yet, $O(N_L)=e^N$ meaning the GLN complexity is also $O(e^N)$.
In contrast, the complexity of the ACN if $O\left((L^2 - N\langle k^2\rangle + N\langle k\rangle) f\right)$ where $f$ is the number of fabrics averaged over.
For example, in a sparse network with $N\sim L=20$, computing the GLN takes approximately the same time as computing the ACN on a network with 20,000 nodes.
Finally, the ACN can be measured on networks without loops and can be analytically estimated via Eq.~\ref{eqn:avg-cross} (see SI \ref{si:gln}).

Also note that degree heterogeneity reduces the computational complexity of the ACN.
To test this prediction, we generated networks with degree distribution drawn from a power law with varying degree exponent $\gamma$, 
finding that small $\gamma$ has lower computational cost (see SI \ref{si:gln}).

In conclusion, we have shown how the entanglement of a physical network is simultaneously affected by the network layout and the network topology.
While network density $\left(N\langle k\rangle\right)^2$ sets the magnitude of entanglement, hubs and community structure reduce its value.

Our results can be further improved by examining how to measure the community structure and estimate $p$ using both the network layout and topology, improving the predictive power of Eq.~\ref{eqn:avg-cross}.
Presently Eq.~\ref{eqn:avg-cross} only offers a bounded range on the possible ACN.
Furthermore, we assumed that $\langle l\rangle^*$, degree heterogeneity, and community structure are independent variables, which is not generally true \cite{barabasi2003scale} (see SI~\ref{si:max-crossings} and \ref{si:link-length}).
Furthermore, understanding the impact of degree heterogeneity and community structure on $\langle l\rangle^*$ will glean better insight into the full role of network topology on the ACN.

\bigskip
We thank Jinha Park for his initial exploration of fabric projections to study non-isotopic layouts of networks.
We also thank Márton Pósfai, Ivan Bonamassa, Csaba Both, and Harrison Hartle for helpful discussions regarding this work.
This research was supported by the NSF award No 2243104 -- COMPASS. ALB is also supported by the European Union's Horizon 2020 research and innovation program No 810115 – DYNASNET.

\bibliography{ref}

\clearpage
\onecolumngrid

\begin{centering}
\textbf{Supplementary Information}\\
\end{centering}

\section{SI I: Graph Linking Number}
\label{si:gln}

The graph linking number measures the amount of linking between cycles in the network \cite{liu2021isotopy}. 
It suffers from high computational complexity and is only applicable to networks with cycles.
For LPNs, the computational complexity of the GLN is $O(N_L^2\langle c\rangle^2)$, where $N_L$ is the number of loops and $\langle c\rangle$ is the average cycle length \cite{liu2021isotopy}. 
As $O(N_L)=e^N$ \cite{bianconi2005loops},
Liu et.~al.~approximate $N_L$ by $s(L-N+1)$ where $s$ is the number of spanning trees used, $N$ is the number of nodes, and $L$ is the number of links. 
In contrast, the proposed ACN is $O(fm_{\max})$ which is at most $O(fL^2)$.
Due to the self-averaging behavior of the ACN, we can assume $f=1$.

\begin{figure}[h!]
    \centering
    \includegraphics[width=\columnwidth]{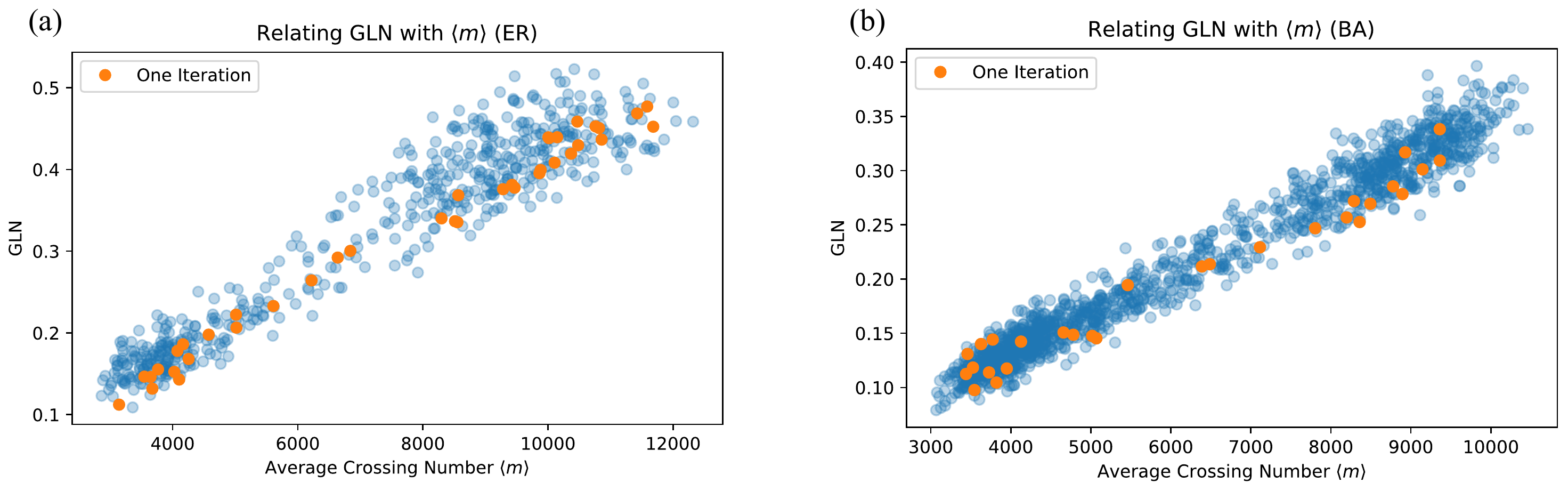}
    \caption{We calculate the ACN and GLN for ER and BA networks at varying energy levels and plot the ACN vs.~GLN. We repeat this process for many iterations and plot all iterations at once. The orange points represent one iteration. The average correlation among iterations is $r=0.99$.}
    \label{fig:gln}
\end{figure}

Additionally, we find that the average crossing number and GLN correlate for random, regular, and heterogeneous topologies (r=0.99) making the ACN a useful substitute for the GLN.
We test this by calculating the GLN and average crossing number for lattice networks, random networks, and BA networks with different embeddings (Fig.~\ref{fig:gln}).
In each case, we observe strong correlation.
However it should be noted that the GLN will not always correlate with the ACN.
For example, in trees the ACN may be very large but regardless of the embedding, the GLN will be zero as there are no loops in the network.

\begin{figure}[h!]
    \centering
    \includegraphics[width=.5\columnwidth]{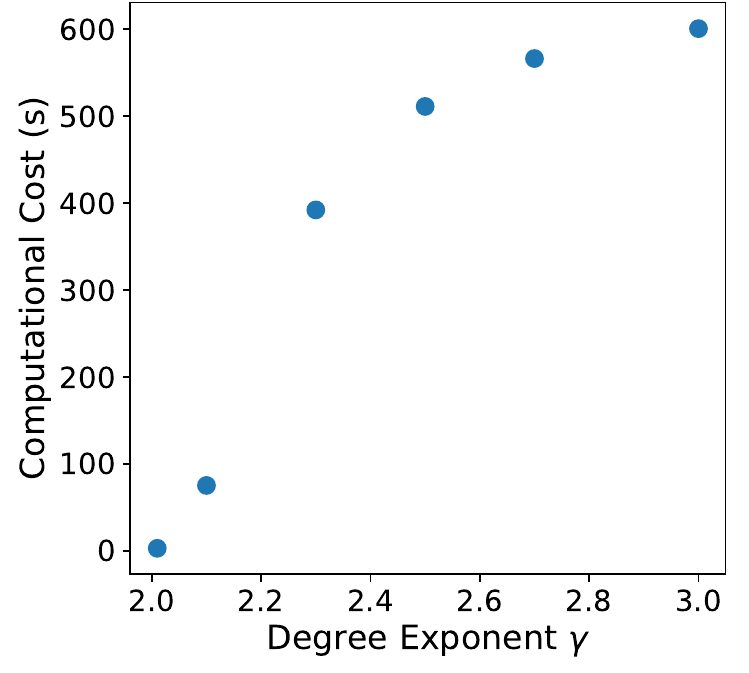}
    \caption{We generate networks with fixed average degree using a hypersoft configuration model where the degree distributions are drawn from a Pareto distribution with varying degree exponents $\gamma$.
    For each $\gamma$, we measure the ACN and calculate the computational cost in seconds.
    There is a clear trend that as $\gamma$ increases, so does the computational cost.}
    \label{fig:cost}
\end{figure}

Because of the relationship between the second moment and $m_{\max}$, we expect the second moment to reduce the computational cost of calculating the ACN.
To test this, we generate 20 networks using a hypersoft configuration model with degree exponent $\gamma=2.01,2.1,2.3,2.5,2.7$ as well as a BA model where $\gamma=3$.
Each network has $N=10^3$ and its expected average degree is $\langle k\rangle=6$.
Averaging over all networks with the same degree exponent, we find that as $\gamma$ decreases, so does the computational cost (see Fig.~\ref{fig:cost}).
Notably, it is near zero for $\gamma=2.01$ (2.86 s).

\newpage

\section{SI II: Self-Average Property}
\label{si:self-averaging}

\begin{figure}[h!]
    \centering
    \includegraphics[width=\columnwidth]{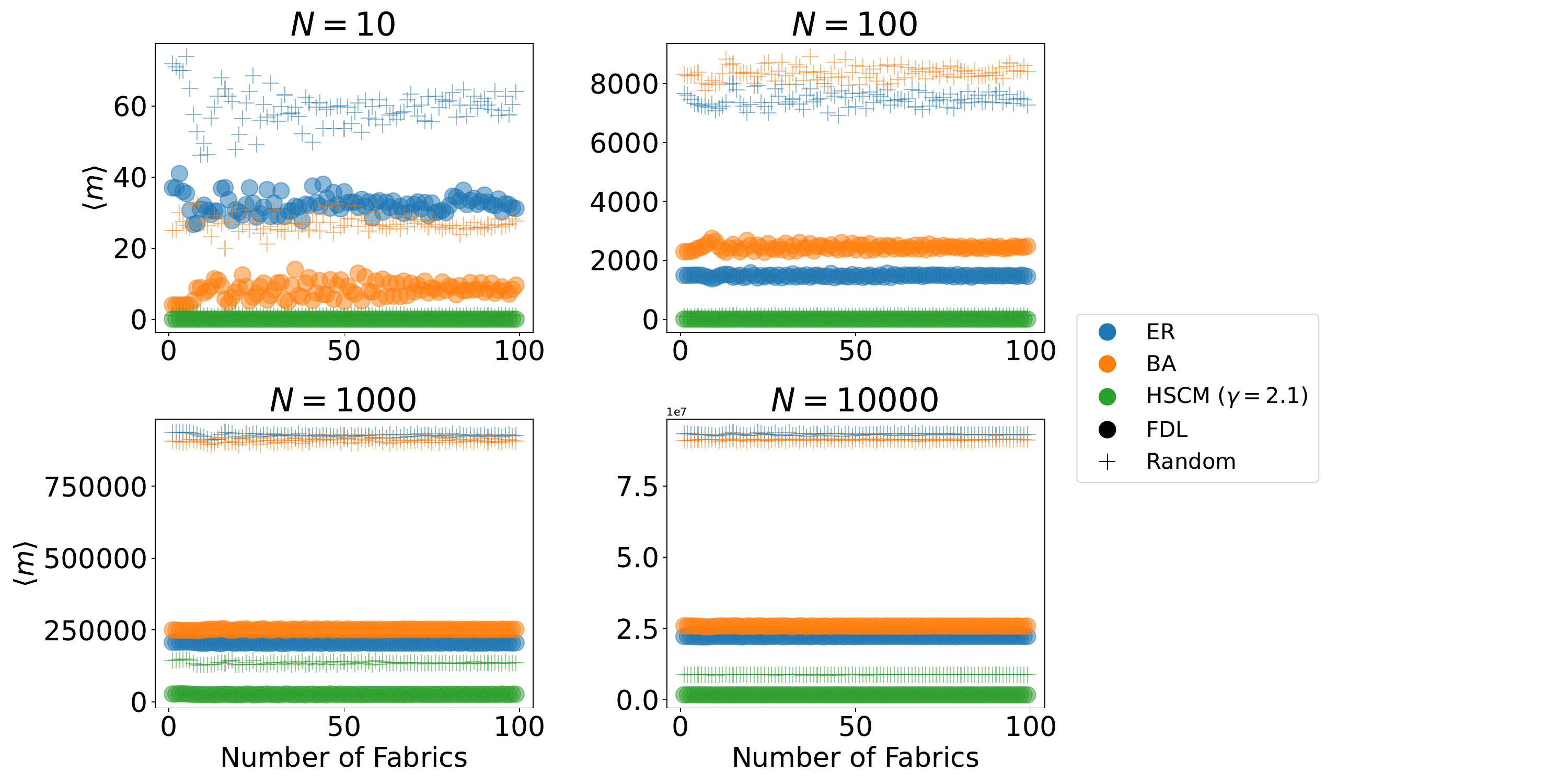}
    \caption{We estimate the ACN for $G(N,p)$, BA and configuration model networks with $P(k)\sim k^{-2.1}$ using different numbers of fabrics for both force-directed and random layouts. The plot shows the number of fabrics used, $F$, versus the ACN $\langle m\rangle$.}
    \label{fig:self-averaging}
\end{figure}

To test whether the ACN exhibits self-averaging behavior, we generate networks $(N=10^1,10^2,10^3,10^4)$ using an $G(N,p)$ model, BA model, and hypercanonical configuration model with degree distribution drawn from a power law with $\gamma=2.1$.
Each network is first embedded optimally using an accelerated force-directed layout algorithm \cite{both2023accelerating} and then embedded randomly in a sphere ($r=N^{1/3}$).
We estimate the ACN for each network by projecting $F$ fabrics for $F\in[1,100]$.
In Fig.~\ref{fig:self-averaging}, we see that as $N\rightarrow\infty$ we merely need one fabric to estimate the ACN.

\newpage
\section{SI III: Estimating the Number of Crossings}
\label{si:max-crossings}

The total number of possible crossings is determined by the pairs of links that can cross.
Links can only cross if they are not connected to the same node.
Let $N$ be the number of nodes, $L$ the number of links, $V(G)$ the set of nodes and $k_i$ the degree of node $i$.
The maximum number of possible crossings is bounded above by
\begin{equation}
    m_{\max}\leq\binom{L}{2}-\sum_{i\in V(G)}\binom{k_i}{2},
    \label{eqn:upper-bound}
\end{equation}
where the first term counts the number of potential link pairs and the second term removes link pairs connected to the same node.
Furthermore, according to \cite{schaefer2012graph}, $\langle m\rangle_{\max}$ is bounded below by 
\begin{equation}
    m_{\max}\geq\frac{1}{3}\left(\binom{L}{2}-\sum_{i\in V(G)}\binom{k_i}{2}\right)
    \label{eqn:lower-bound}
\end{equation}
so we expect that
\begin{equation}
    m_{\max}\sim\binom{L}{2}-\sum_{i\in V(G)}\binom{k_i}{2}.
    \label{eqn:max-scaling}
\end{equation}
Expanding Eq.~\ref{eqn:max-scaling} we have
\begin{align}
    m_{\max}&\sim\frac{L(L-1)}{2}-\sum_{i\in V(G)}\frac{k_i(k_i-1)}{2}\\
    &=\frac{L(L-1)}{2}-\sum_{i\in V(G)}\left(\frac{k_i^2}{2}-\frac{k_i}{2}\right)\\
    &=\frac{L(L-1)}{2}-\sum_{i\in V(G)}\frac{k_i^2}{2}+\sum_{i\in V(G)}\frac{k_i}{2}\\
    &=\frac{L(L-1)}{2}-\frac{N}{2}\langle k^2\rangle+\frac{N}{2}\langle k\rangle
    \label{eqn:max-scaling-expanded}
\end{align}
where $\langle k\rangle=\frac{1}{N}\sum_{i\in V(G)}k_i$ and $\langle k^2\rangle=\frac{1}{N}\sum_{i\in V(G)}k_i^2$ are the first and second moments of the degree distribution $P(k)$, respectively.
For large $L$, we assume that $L(L-1)\approx L^2$.
Thus we can approximate Eq.~\ref{eqn:max-scaling-expanded} with
\begin{equation}
    m_{\max}\sim\frac{1}{2}\left(L^2-N\langle k^2\rangle+N\langle k\rangle\right).
    \label{eqn:max-scaling-approx}
\end{equation}
With this estimate, we can now calculate the ACN $\langle m\rangle$.
We assume that the ACN is merely $\langle m\rangle=m_{\max}P(\mathcal{E})$ where $P(\mathcal{E})$ is the probability that a random link pair crosses in embedding $\mathcal{E}$.
Next we will derive $P(\mathcal{E})$ for both random and optimal embeddings.

\emph{Random Layouts} -- For networks embedded randomly in a ball, we can calculate $P(\mathcal{E}_{rand})$ using Sylvester's Four Point Theorem.
This theorem states that the probability any randomly placed four points in a disk create a convex quadrilateral is $1-\frac{35}{12\pi^2}$ \cite{pfiefer1989historical,watson1865question,woolhouse1867some}.
Consequently, the probability that two randomly placed line segments in a disk cross is $\frac{1}{3}\left(1-\frac{35}{12\pi^2}\right)$.
By randomly embedding a network and taking many projections, we can assume that the network will have its links approximately randomly placed in a disk in each fabric, though not fully random.
Consequently, for sparse networks where $L\sim N$, we expect
\begin{equation}
    \langle m\rangle_{rand}\approx\frac{1}{24}\left(1-\frac{35}{12\pi^2}\right)\left(N\langle k\rangle\right)^2\left(1-\frac{\langle k^2\rangle}{N\langle k\rangle^2}+\frac{1}{N\langle k\rangle}\right).
    \label{eqn:m-random-layout}
\end{equation}

\emph{Optimal Layouts} - In optimal layouts, not all link pairs have an equal probability of crossing in a fabric.
In fact, force-directed layouts will separate nodes based on their community structure \cite{both2023accelerating}.
To take this into account, we assume that all links within the same community have an equal likelihood of crossing each other and that links in distinct communities will not cross.

To calculate $\langle m\rangle_{opt}$ for optimal layouts, we must adjust $m_{\max}$ by first identifying how many links are within communities and how many are inter-community links.
Let $p$ be the probability that a link is an inter-community link, i.e.~a link between distinct communities.
Then the number of possible link pairs within one community is $\binom{\frac{(1-p)L}{C}}{2}$.
Similarly, we expect $pL$ inter-community links.
These inter-community links have the potential to cross any of the other $L-1$ links in the network.
Consequently, they will contribute
\begin{align}
    \binom{pL}{2} + pL\left(L-pL\right)&\approx\frac{p^2L^2}{2}+pL^2-p^2L^2\\
    &=pL^2 - \frac{p^2L^2}{2}
\end{align}
possible crossings to the fabric.
If $p$ is small, then we can assume the inter-community links contribute to $pL^2$ potential crossings.
Combining these two results, the maximum number of crossings in an optimal layout becomes
\begin{align}
    m_{\max}^{opt}&\sim C\binom{\frac{(1-p)L}{C}}{2}+pL^2-\sum_{i\in V(G)}\binom{k_i}{2}\\
    &=C\left(\frac{(1-p)^2L^2}{2C^2}\right)+pL^2-\frac{N\langle k^2\rangle}{2}+\frac{N\langle k\rangle}{2}\\
    &\sim\frac{(1-p)^2L^2}{C}+2pL^2-N\langle k^2\rangle+N\langle k\rangle.
    \label{eqn:m-max-opt}
\end{align}

If we assume that the communities are randomly arranged in space and that each community resembles a random embedding in its subnetwork, then we can simply use the coefficient we found for random layouts to estimate $P(\mathcal{E}_{opt})$.
However, we must modify this coefficient by the normalized average link length $\langle l\rangle^*$.
This is because networks with long links have more crossings.
In random layouts, all links are statistically equivalent in length.
For optimal layouts this may not be true.
Thus we have $P(\mathcal{E}_{opt})=\langle l\rangle^*P(\mathcal{E}_{rand})$.
We can then combine $P(\mathcal{E}_{opt})$ with Eq.~\ref{eqn:m-max-opt} to get
\begin{align}
    \langle m\rangle\approx\frac{1}{24}\Bigg(1-\frac{35}{12\pi^2}\Bigg)\langle l\rangle^*\Bigg(\frac{(1-p)^2}{C}+2p-\frac{\langle k^2\rangle}{N\langle k\rangle^2}+\frac{1}{N\langle k\rangle}\Bigg).
    \label{eqn:avg-cross-fdl}
\end{align}
Note that as $\langle l\rangle^*\rightarrow 1$ and $p\rightarrow 0$, Eq.~\ref{eqn:avg-cross-fdl} reduces to Eq.~\ref{eqn:m-random-layout}.

The assumption that $P(\mathcal{E}_{opt})$ scales with $\langle l\rangle^*$ comes from \cite{liu2021isotopy}, where it is shown that the network's elastic energy relates linearly with the graph linking number.
In SI \ref{si:gln} we show that the GLN and the ACN are highly correlated.
Consequently, we expect that $\langle m\rangle\sim f(\langle l\rangle)^*)$ where $f$ is some linear transformation of $\langle l\rangle^*$ and we assume that $f(\langle l\rangle^*)\sim \langle l\rangle^*P(\mathcal{E})$.

\begin{figure}[h!]
    \centering
    \includegraphics[width=\columnwidth]{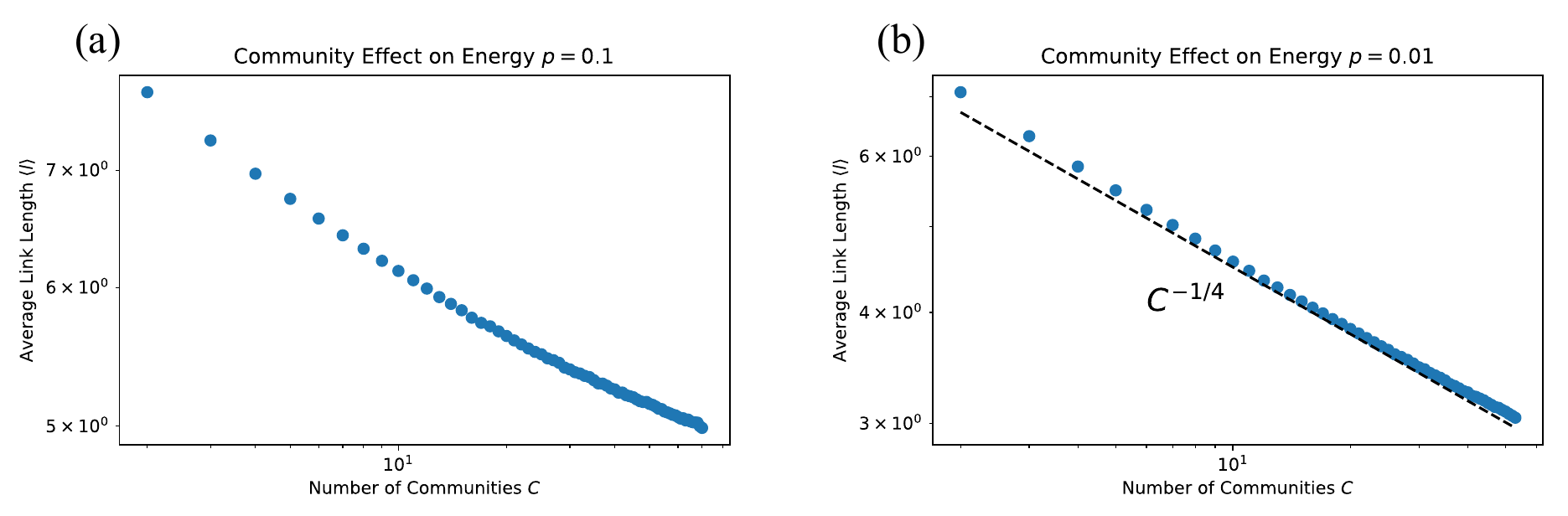}
    \caption{(a) The average link length as a function of the number of communities for a network with $p=0.1$. We see a decreasing trend. (b) The average link length as a function of the number of communities for a network with $p=0.01$. Empirically we see this follows a power-law $C^{-1/4}$.}
    \label{fig:avg-link-C}
\end{figure}

In order to arrive at this approximation we made various assumptions which do not always hold.
First, we assumed that communities are homogeneous in size.
If our network has heterogeneous communities, Equation \ref{eqn:avg-cross} will not simplify into only terms of the number of communities.
Second, we assume that all links within each community are randomly embedded within the community.
This may not be true as the community may not have a random link structure.
We also assumed that our communities are distinct from each other and that each node is clearly within one community.
In real networks, there are overlapping communities \cite{fortunato2010community,palla2005uncovering}.
Next, we assume that the average link length of the network affects the probability of links interacting as a multiplicative factor.
Lastly, we assume that $\langle k^2\rangle$, $C$, and $\langle l\rangle^*$ are all independent.
This has yet to be shown analytically and may not hold for all networks.
In fact, measurements indicate the average link length may depend on $C$ (see Fig.~\ref{fig:avg-link-C}).

\newpage

\section{SI IV: Simulated Annealing Pipeline}
\label{si:mcmc}

\begin{figure}[h!]
    \centering
    \includegraphics[width=\columnwidth]{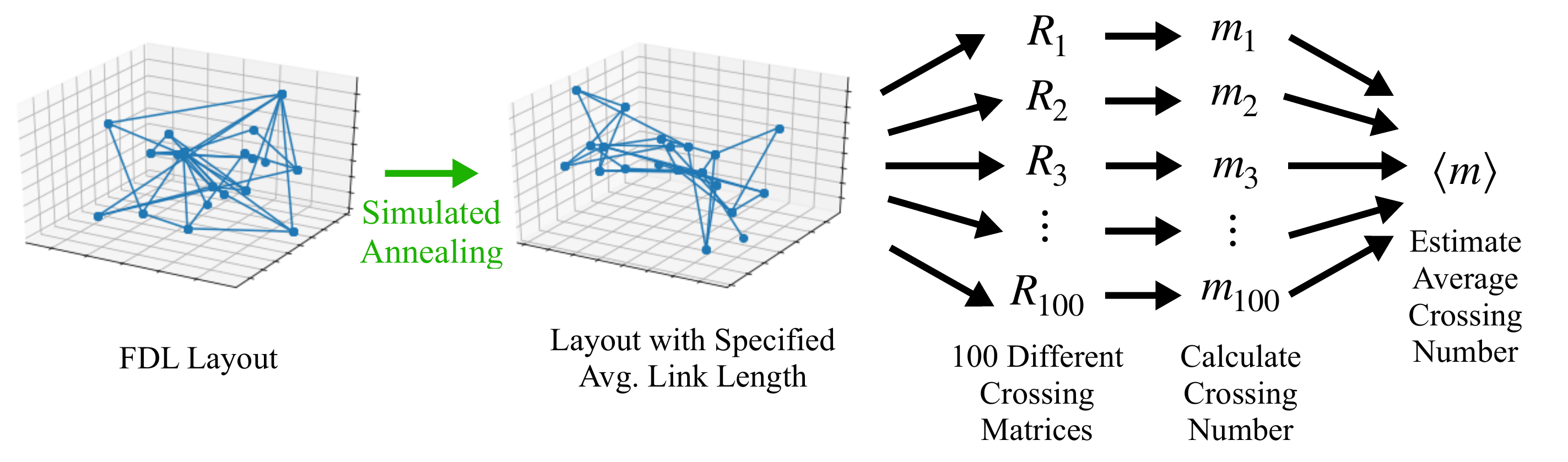}
    \caption{Simulated Annealing Pipeline}
    \label{fig:mcmc}
\end{figure}

To generate networks with a specified average link length $\langle l\rangle$, we generate a force directed layout of a network using NeuLay-2 \cite{both2023accelerating}.
Then at each time step, we randomly choose a node and alter its position by selecting a new position within some $\epsilon$ distance using a normal distribution.
We then calculate the new average link length.
Using simulated annealing, we continue this process until we arrive at a network embedding with a predefined average link length.
This network is then projected many times to calculate the ACN (see Fig.~\ref{fig:mcmc}).

\newpage 

\section{SI V: Second Moment Relationship With Energy In Force-Directed Layouts}
\label{si:link-length}

In force directed layouts, hubs are pulled to the center of the layout in order to reduce energy.
If the hub is sufficiently large, it's links will stretch from the center of the network to all other points in the layout.
Furthermore, a node randomly placed in a ball of radius $N^{1/3}$ will be at a distance $\frac{3}{4}N^{1/3}$ from the center of the ball on average.
Combining these two facts, if the hub conglomerates the majority of the links, then the average link length will simply be $\langle l\rangle=\frac{3}{4}N^{1/3}$.

\begin{figure}[h!]
    \centering
    \includegraphics[width=.7\columnwidth]{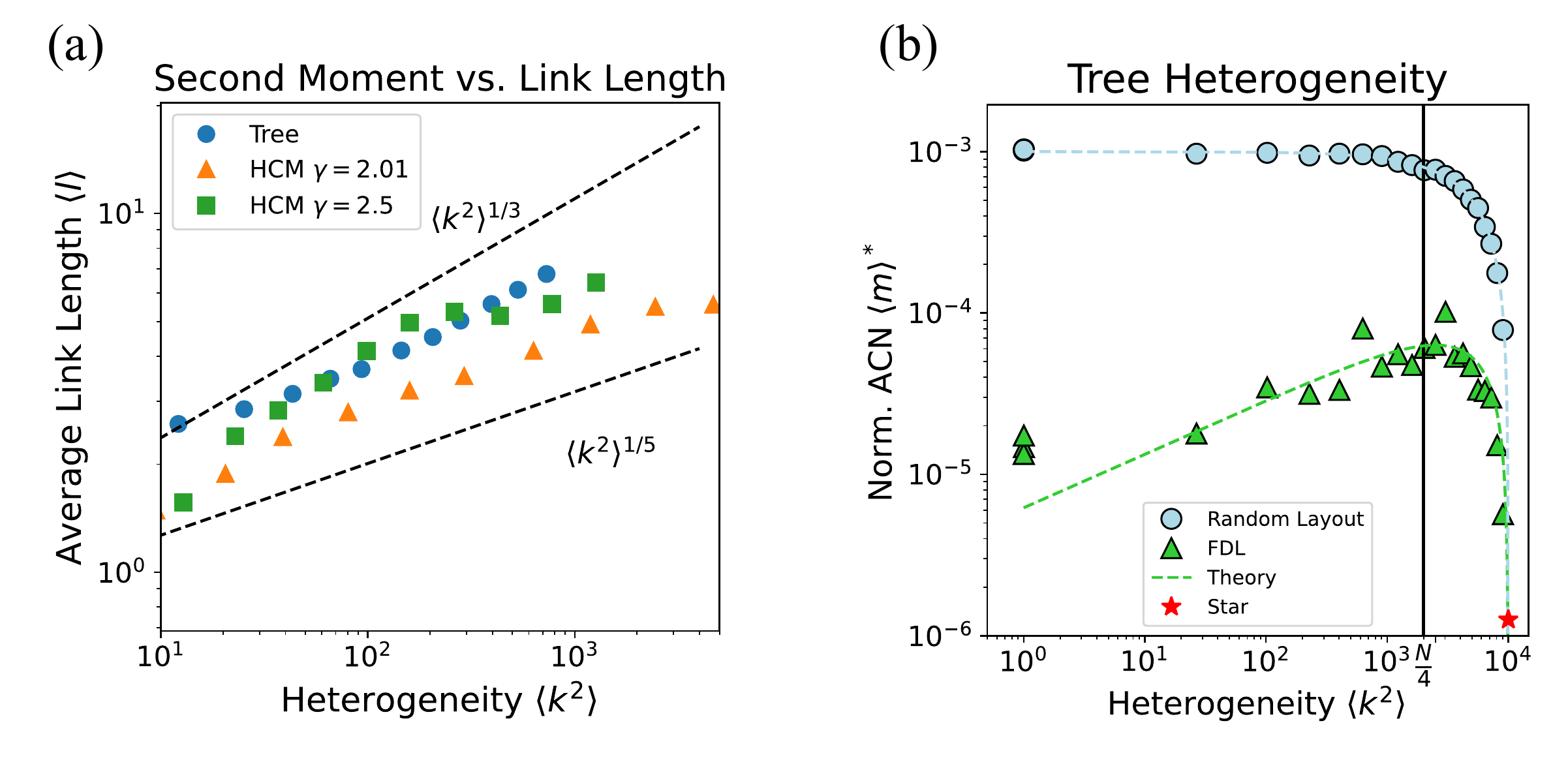}
    \caption{(a) We generate tree networks with varying second moments and two hypercanonical configuration models with power-law degree distributions ($N=10^3$).
    We plot their average link length as a function of network heterogeneity, capturing the predicted dependence on the second moment. 
    (b) We generate a random tree $(N=10^4)$ and slowly rewire each edge to increase the heterogeneity. The tree is embedded randomly and optimally. We show the normalized ACN for each embedding along with the analytical prediction. In the optimal layout, $\langle l\rangle^*$ is substituted with $\langle k^2\rangle^{1/3}$ in the analytical prediction and the maximum ACN is marked at $\frac{N}{4}$.}
    \label{fig:tree}
\end{figure}

For a tree, $\langle k^2\rangle\sim N$.
Thus the average length of a link connected to the hub will be $\frac{3}{4}\langle k^2\rangle^{1/3}$, obtaining $\langle l\rangle^*\sim\langle k^2\rangle^{1/3}$.
We see similar behavior for power-law configuration models with sufficiently low $\gamma$ (see Fig.~\ref{fig:tree}(a)).

Let us assume that the normalized average link length of a force-directed layout scales with the second moment to some power, as observed in trees.
Let $\tau$ be the strength of the dependence $\langle l\rangle^*$ has on the second moment,
giving $\langle l\rangle^*\sim\langle k^2\rangle^\tau$.
In the case of a tree, $\tau=\frac{1}{3}$.
Thus as $\langle k^2\rangle\rightarrow\infty$, we have $\langle l\rangle\rightarrow1$, thereby increasing the ACN.

However, if $\langle k^2\rangle$ is sufficiently large, it will reduce the ACN (see Fig.~\ref{fig:combined}(b)).
By substituting $\langle k^2\rangle^\tau$ for $\langle l\rangle$, we can solve for the critical point where $\langle k^2\rangle$ is large enough to reduce the number of crossings:
\begin{equation}
    \langle m\rangle\sim\langle k^2\rangle^{\tau}\Bigg(\frac{(1-p)^2}{C^{-1}}+2p-\frac{\langle k^2\rangle}{N\langle k\rangle^2}+\frac{1}{N\langle k\rangle}\Bigg).
\end{equation}

By taking the derivative with respect to $\langle k^2\rangle$, we find
\begin{align}
    \frac{\partial\langle m\rangle_{FDL}}{\partial\langle k^2\rangle}=\tau\langle k^2\rangle^{\tau-1}\Bigg(\frac{(1-p)^2}{C^{-1}}+2p-\frac{\langle k^2\rangle}{N\langle k\rangle^2}+\frac{1}{N\langle k\rangle}\Bigg)-\frac{\langle k^2\rangle^\tau}{N\langle k\rangle^2}.
\end{align}
Setting this equal to 0 and dividing by $\langle k^2\rangle^{\tau-1}$ we find that
\begin{equation}
    \frac{\langle k^2\rangle}{N\langle k\rangle^2}=\tau\Bigg(\frac{(1-p)^2}{C^{-1}}+2p-\frac{\langle k^2\rangle}{N\langle k\rangle^2}+\frac{1}{N\langle k\rangle}\Bigg)
\end{equation}
\begin{equation}
    \langle k^2\rangle\left(\frac{\tau+1}{N\langle k\rangle^2}\right)=\tau\Bigg(\frac{(1-p)^2}{C^{-1}}+2p+\frac{1}{N\langle k\rangle}\Bigg)\end{equation}
\begin{equation}
    \langle k^2\rangle=\frac{\tau N\langle k\rangle^2}{\tau+1}\left(\frac{(1-p)^2}{C^{-1}}+2p+\frac{1}{N\langle k\rangle}\right)\end{equation}
\begin{equation}
    \rightarrow\frac{\tau N}{(\tau+1)}\left(\frac{(1-p)^2}{C^{-1}}+2p\right)
\end{equation}
as $N\rightarrow\infty$ and $\langle k\rangle$ is constant.
We apply a change of variables $\tau=\frac{1}{\rho}$ as we expect $\tau<1$.
Then we have
\begin{align}
    \frac{\tau}{\tau+1}=\left(\frac{1}{\rho}\right)\left(\frac{\rho}{\rho+1}\right)=\frac{1}{\rho+1}
\end{align}
implying hubs reduce entanglement only if $\langle k^2\rangle>\frac{N}{(\rho+1)}\left(\frac{(1-p)^2}{C^{-1}}+2p\right)$.

We tested our prediction using trees.
We choose trees because we analytically have $\tau=\frac{1}{3}$ for $p=1$.
We generate a random tree and slowly rewire each edge to some hub.
As we do so, we calculate the ACN for a force directed layout and random layout. 
To test the accuracy of our experiment, we plot our analytical predictions of the ACN using Equation \ref{eqn:avg-cross-fdl} and substitute $\langle k^2\rangle^{1/3}$ for $\langle l\rangle^*$ for the force-directed layout prediction.
In Fig.~\ref{fig:tree}, we see that our experiments match with our analytical expectation.
Additionally, we find that the heterogeneity of the trees reduces the ACN when $\langle k^2\rangle>\frac{N}{4}$, in line with our measurements (see Fig.~\ref{fig:tree}(b)).

\newpage

\section{SI VI: Estimating the Scaling of $\langle m\rangle$ for Network Models}
\label{si:estimating-scaling-behvaior}

\emph{Regular Network} -- In a regular network, all nodes have degree $k$.
Consequently, we have $\langle k\rangle=k$ and $\langle k^2\rangle=k^2$.
In this case, (\ref{eqn:avg-cross-fdl}) becomes
\begin{equation}
    \langle m\rangle\sim\left(Nk\right)^2\langle l\rangle^*\Bigg(\frac{(1-p)^2}{C^{-1}}+2p-\frac{k^2}{Nk^2}+\frac{1}{Nk}\Bigg)\end{equation}
\begin{equation}
    =\left(Nk\right)^2\langle l\rangle^*\left(\frac{(1-p)^2}{C^{-1}}+2p-\frac{1}{N}+\frac{1}{Nk}\right).
\end{equation}

\emph{Random Networks (Erd\H{o}s-Rényi)} -- In a random network we have $\langle k^2\rangle=\langle k\rangle(1+\langle k\rangle)$.
Thus (\ref{eqn:avg-cross-fdl}) becomes
\begin{equation}
    \langle m\rangle\sim\left(N\langle k\rangle\right)^2\langle l\rangle^*\Bigg(\frac{(1-p)^2}{C^{-1}}+2p-\frac{\langle k\rangle\left(1+\langle k\rangle\right)}{N\langle k\rangle^2}+\frac{1}{N\langle k\rangle}\Bigg)\end{equation}
    \begin{equation}=\left(N\langle k\rangle\right)^2\langle l\rangle^*\Bigg(\frac{(1-p)^2}{C^{-1}}+2p-\frac{1+\langle k\rangle}{N\langle k\rangle}+\frac{1}{N\langle k\rangle}\Bigg)\end{equation}
    \begin{equation}=\left(N\langle k\rangle\right)^2\langle l\rangle^*\left(\frac{(1-p)^2}{C^{-1}}+2p-\frac{1}{N}\right).
\end{equation}

\emph{BA Networks} -- In a BA network, we have $\langle k^2\rangle\sim \log N$.
Thus (\ref{eqn:avg-cross-fdl}) becomes
\begin{equation}
    \langle m\rangle\sim\left(N\langle k\rangle\right)^2\langle l\rangle^*\Bigg(\frac{(1-p)^2}{C^{-1}}+2p-\frac{\log N}{N\langle k\rangle^2}+\frac{1}{N\langle k\rangle}\Bigg).
\end{equation}

\emph{Scale-Free Network $2<\gamma<3$} -- In a scale free network we expect that $\langle k^2\rangle\sim N^{(3-\gamma)/(\gamma-1)}$.
In this case, (\ref{eqn:avg-cross-fdl}) becomes
\begin{equation}
    \langle m\rangle\sim\left(N\langle k\rangle\right)^2\langle l\rangle^*\Bigg(\frac{(1-p)^2}{C^{-1}}+2p-\frac{N^{\frac{3-\gamma}{\gamma-1}}}{N\langle k\rangle^2}+\frac{1}{N\langle k\rangle}\Bigg)\end{equation}
    \begin{equation}=\left(N\langle k\rangle\right)^2\langle l\rangle^*\Bigg(\frac{(1-p)^2}{C^{-1}}+2p-\frac{N^{\frac{4-2\gamma}{\gamma-1}}}{\langle k\rangle^2}+\frac{1}{N\langle k\rangle}\Bigg).
\end{equation}

\emph{Scale-free Network $1<\gamma<2$} -- In a scale free network with diverging first and second moment, we have that $\langle k\rangle\sim N^{(2-\gamma)/\gamma}$ and $\langle k^2\rangle\sim N^{\gamma(1-\gamma)/2}$ \cite{seyed2006scale}.
In this case we have that $\frac{\langle k^2\rangle}{\langle k\rangle^2}\sim N^{\frac{1}{2}\gamma^2(1-\gamma)/(\gamma-2)}.$
With this (\ref{eqn:avg-cross-fdl}) becomes
\begin{equation}
    \langle m\rangle\sim N^{4/\gamma}\langle l\rangle^*\Bigg(\frac{(1-p)^2}{C^{-1}}+2p-N^{\frac{1}{2}\gamma^2(1-\gamma)/(\gamma-2)-1}+N^{-2/\gamma}\Bigg).
\end{equation}

\newpage

\section{SI VII -- Physical Network Data}
\label{si:data}

We collected physical network statistics describing various data systems including coral structures, an arabidopsis plants \cite{haolin}, tropical trees \cite{gonzalez2018estimation}, parts of the vascular system in the human lung, and three neurons (human, fly from the hemibrain, and fly from the male adult nerve cord (MANC)). 
Each dataset contains multiple networks.
We present the average network statistics for each network in Table \ref{tab:data-stats}.

\begin{table}
    \scriptsize
    \begin{tabular}{c|c|c|c|c|c|c}
          & $N$ & $\langle k\rangle$ & $\langle k^2\rangle$ & $\langle l\rangle^*$ & $C$ & $\langle m\rangle$\\
          \hline\hline
        Coral & 4677.39 & 2&4.03 & 0.01 & 65.89 & $1.2\times10^5$ \\
        \hline
        Trop. Tree & 4474.59 &2& 4.62 & 0.03 & 64.59 & $3.2\times 10^4$ \\
        \hline
        Arad. Plant & 1371.64 &2& 3.95 & 0.02 &  59.3 & $5.4\times 10^3$\\
        \hline 
        Lung Vascular & 3295.94 &2&4.04 & 0.01 & 56.29 & $2.6\times 10^4$ \\
        \hline Neuron (H) & 1451.9 &2& 4.08 & 0.03 & 41.93 & $3.2\times 10^3$ \\
        \hline
        \begin{tabular}{@{}c@{}}Hemibrain \\Neuron (F) \end{tabular}& 3216.38 & 2&4.15 & $0.01$ & 50.8 & $3.6\times 10^{3}$ \\
        \hline \begin{tabular}{@{}c@{}}MANC \\Neuron (F) \end{tabular}& 737.96 & 2&4.11 & 0.02 & 25.8 & $3.8\times 10^1$ 
    \end{tabular}
    \caption{Physical network statistics of coral structures, arabidopsis plants, parts of the vascular system of the human lung, and three neurons (human, fruit fly from the hemibrain, and fruit fly from the male adult nerve cord (MANC)).}
    \label{tab:data-stats}
\end{table}

\newpage




\newpage
\section{SI VIII: Normalized ACN of Real Networks}
\label{si:normalized-acn}

\begin{figure}[h]
    \centering
    \includegraphics[width=\columnwidth]{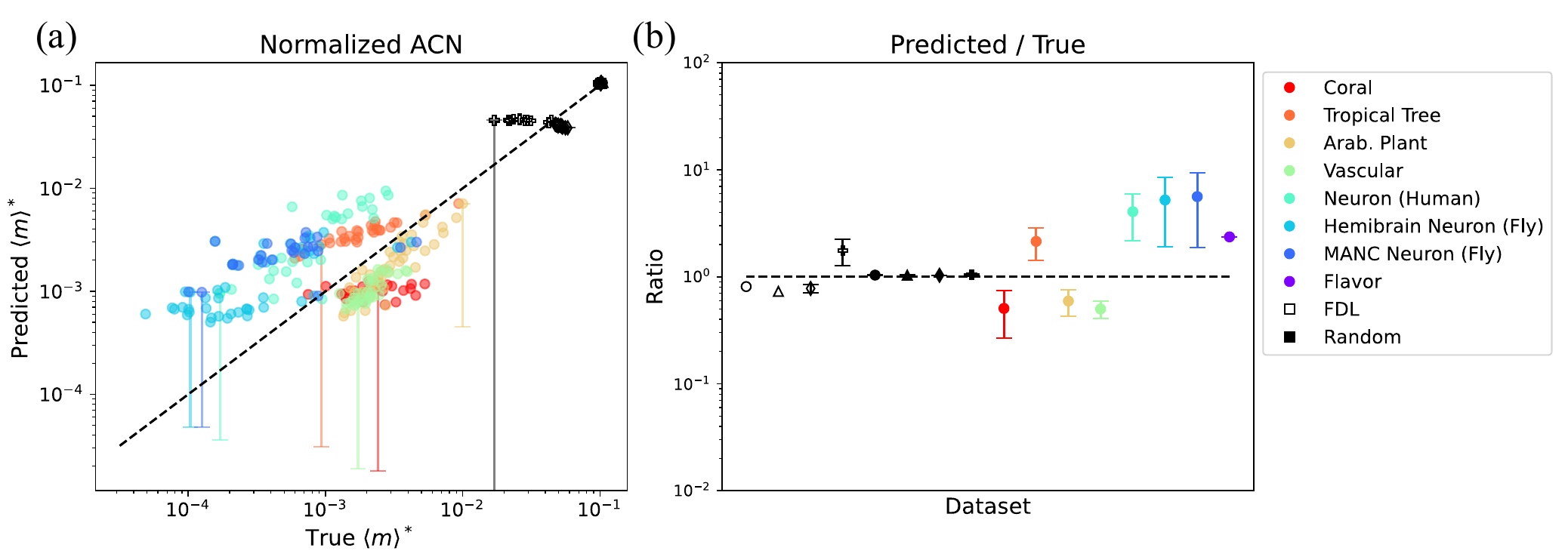}
    \caption{(a) The true normalized ACN vs.~the predicted normalized ACN for synthetic network models and real datasets.
    The black line indicates $x=x$.
    For each dataset, we compute the normalized ACN for $p=0$ and $p=1$, represented by the errorbar (one for each dataset).
    Each dataset point is the prediction for $p=1$.
    (b) The ratio between the predicted ACN and the true ACN of each dataset.}
    \label{fig:normalized-acn}
\end{figure}

The leading term of the ACN is the network density.
To confirm that our prediction is not merely capturing the effect of network density, we measure the normalized ACN $\langle m\rangle^*=\frac{\langle m\rangle}{L^2}$.
Consequently, our predictions of $\langle m\rangle^*$ only depend on the normalized average link length $\langle l\rangle^*$ and the maximum number of crossings $m_{\max}$.
In Fig.~\ref{fig:normalized-acn}(a), we see that we accurately predict the true normalized ACN for both network models and real physical networks.
In the normalized ACN, the need to consider community structure in the neurons, tropical trees, flavor network, and configuration network models is emphasized.

In Fig.~\ref{fig:normalized-acn}(b), we show the ratio between our prediction of the ACN and the true ACN for $p=1$ and we consistently see ratio's close to one.
The network's with ratio larger than one are the networks where community structure should be taken into account when predicting $\langle m\rangle$.

\end{document}